\documentclass[preprint,amsmath,amssymb,aps,pre,longbibliography]{revtex4-1}

\usepackage{graphicx}
\usepackage{dcolumn}
\usepackage{bm}
\usepackage{color}
\usepackage[utf8]{inputenc}
\usepackage{cancel}

\newcommand{\vX}{\vec{X}}
\newcommand{\vY}{\vec{Y}}

\begin{document}

\title{Physical meaning of nonextensive term in Massieu functions} 

\author{M. Hoyuelos}
\email{hoyuelos@mdp.edu.ar}
\affiliation{Instituto de Investigaciones F\'isicas de Mar del Plata (IFIMAR-CONICET), Departamento de F\'isica, Facultad de Ciencias Exactas y Naturales, Universidad Nacional de Mar del Plata, Funes 3350, 7600 Mar del Plata, Argentina}
\author{M. A. Di Muro}
\affiliation{Instituto de Investigaciones F\'isicas de Mar del Plata (IFIMAR-CONICET), Departamento de F\'isica, Facultad de Ciencias Exactas y Naturales, Universidad Nacional de Mar del Plata, Funes 3350, 7600 Mar del Plata, Argentina}
\author{P. Giménez}
\affiliation{Departamento de Matemática, Facultad de Ciencias Exactas y Naturales, Universidad Nacional de Mar del Plata, Funes 3350, 7600 Mar del Plata, Argentina}

\begin{abstract}
In this paper we explore the significance of nonextensive terms in Massieu functions. Finite-size effects are in many cases dominated by a term proportional to the surface area. Nevertheless, in numerical simulations of finite systems with periodic boundary conditions, the nonextensive term can become the dominant correction to Massieu functions. This paper presents a general approach linking these nonextensive terms to thermodynamic fluctuations, demonstrating that equations of state inherently encode this information. Numerical simulations corroborate our results. The examples used are hard sphere and hard disk fluids and a one-dimensional spin lattice, emphasizing the applicability of the results across different classes of systems.
\end{abstract}


\maketitle

\section{Introduction}

Nonextensive terms in the Boltzmann definition of entropy are usually neglected in the thermodynamic limit, where the results of statistical mechanics converge to those of thermodynamics. The differences between extensive quantities of thermodynamic and statistical mechanics are of the order $\ln N$, where $N$ is the number of elements that compose the system. The logarithmic relationship stems from the number of possible permutations of $N$ elements: $N!$. When we take the logarithm of $N!$, it includes a term proportional to $\ln N$. This term can become significant, especially in molecular simulations of finite systems employing periodic boundary conditions \cite{landaubinder}; in the examples of small systems analyzed here, this term has a contribution between 4\% and 10\% for excess quantities (relative to the ideal value).

We present a general approach based on Massieu functions to demonstrate that these differences can be written in terms of thermodynamic fluctuations that, in turn, can be obtained from thermodynamic quantities. This means that the thermodynamic equation of state contains the information needed to calculate nonextensive terms of statistical mechanics quantities. Or, the other way around, nonextensive terms of statistical mechanics quantities have information about thermodynamic fluctuations. 

Let us call $X_\text{finite}$ a Massieu function or a thermodynamic potential of a finite system and $X$ its value at the thermodynamic limit; $X_\text{finite}$ includes other corrections in addition to the logarithmic one. The most important is proportional to the surface area of the system, $L^2 \sim V^{2/3} \sim N^{2/3}$, where $L$ is the system size, $V$ is the system volume, and $N$ is the particle number; see, for example, Sec. 5.1 in \cite{bedeaux}. Then,
\begin{equation}
    X_\text{finite} - X = O(\ln N) + O(N^{2/3}).\label{e.fin}
\end{equation}
(Additional possible corrections, of order $N^{1/3}$ and $N^{0}$, that correspond to the effects of edges and corners, are not considered here.) In order to determine which of the two corrections in \eqref{e.fin} is dominant, we have to compare the system size with the correlation length of fluctuations \cite{lebowitz-percus}. For a real system with strong interactions that produce a long correlation length similar to the system size, the correction $O(\ln N)$ can be neglected with respect to surface effects; this is the situation studied in the theory introduced by Hill \cite{hill2,bedeaux}. Periodic boundary conditions are often used to simulate bulk properties, but despite the absence of physical surfaces, finite-size effects can still contribute corrections dominant respect to the term $O(\ln N)$; they are usually called \textit{implicit} or \textit{anomalous finite-size effects} \cite{roman}. Salacuse {\it et al}.\ \cite{salacuse1} demonstrated that if the pair distribution function decays rapidly with distance, then implicit finite-size corrections can be neglected, a criterion equivalent to considering the absence of long-range correlations \cite{lebowitz-percus}; in this case the logarithmic correction is dominant. The extreme situation is the ideal system, where interactions can be neglected and the correlation length is zero; for the ideal case with periodic boundary conditions, the only correction is $O(\ln N)$.

Our analysis is restricted to systems with a correlation length smaller than $L$, so implicit finite-size effects can be neglected. The corrections obtained in this case are called \textit{explicit finite-size effects} \cite{roman}. For example, we can analyze the free energy of a dilute gas, with a fixed number of particles $N$, and progressively increase the density by reducing the volume $V$, with periodic boundary conditions. Initially, the system is ideal, interactions can be neglected, and our assumption for the correlation length holds. As the density is increased, interactions become more relevant and the correlation length also increases. A point will be reached for which the correlation length is similar to $L$ and implicit finite-size effects cannot be neglected anymore. At this point, our present analysis does not hold, but there is a range of concentration values, starting from the dilute gas to this point, where our results correctly predict the correction $O(\ln N)$. Actually, for nonideal systems, it is more convenient to analyze excess quantities, that is, for example, the free energy minus its ideal value. It can be seen that, for excess quantities, the correction is $O(N^0)$ instead of $O(\ln N)$.

The excess free energy of a system of particles, calculated using the partition function, includes the nonextensive term mentioned $O(N^0)$, and the excess chemical potential has a correction term $O(N^{-1})$. This term was calculated using the Widom insertion formula in Ref.\ \cite{siepmann}, where the result is presented as a finite-size correction to numerical calculation of the excess chemical potential in a system with periodic boundary conditions. An equivalent expression was derived with a different procedure in Ref.\ \cite{dimuro}, where the nonextensive term was shown to be relevant for the evaluation of transition rates and, in turn, of the diffusion coefficient. In fact, the relationship between self-diffusion and collective diffusion coefficients, known as one of the Darken equations \cite{darken}, was derived in \cite{dimuro} using the correction term for the excess chemical potential. Here we present a general approach, valid for any Massieu function, that includes the mentioned results; that is, it is not restricted to the analysis of the free energy. It is generally more convenient to use Massieu functions instead of thermodynamic potentials to analyze fluctuations (see \cite[Ch.\ 19]{callen}); there are simple relationships that connect them. 

Therefore, we study corrections order $\ln N$ in thermodynamic quantities (Massieu functions or thermodynamic potentials); the procedure cannot be applied to calculate corrections to statistical mechanics quantities involving microscopic information, such as the radial distribution function. Corrections to the structure factor or the radial distribution function of a system of particles were analyzed in Salacuse {\it et al}.\ \cite{salacuse1}, based on the theory developed by Lebowitz and Percus \cite{lebowitz-percus}; it has been applied, for example, to evaluate the structure factor for a model krypton fluid \cite{salacuse2} or fluctuations in the number of hard disks and hard parallel squares deposited on a finite surface \cite{roman2}. The main feature of the present approach is the possibility to obtain a general result for the nonextensive term of any Massieu function without using the radial distribution function. We demonstrate that the nonextensive term is directly related to the determinant of the equilibrium fluctuation matrix; see Eq.\ \eqref{difPsiex} below. We illustrate the generality of the approach applying the results to particle systems and to a spin lattice.

In the next sections we present the basic equations and notation (Sec.\ \ref{s.basic}), the calculation of the difference between thermodynamic and statistical mechanics quantities (Sec.\ \ref{s.nonext}), applications of the results to a system of particles and a spin lattice (Secs.\ \ref{s.particles} and \ref{s.spin}), and the conclusions (Sec.\ \ref{s.conclusions}).

\section{Basic equations and notation}
\label{s.basic}

Let us consider a system described by the extensive variables $\vX = (X_1,X_2,\dots)$. The typical example is $\vX = (U,V,N)$, where $U$, $V$, and $N$ are the internal energy, volume, and particle number. The Boltzmann entropy is
\begin{equation}\label{e.boltzmann}
	\tilde{S}(\vX) = \ln \Omega(\vX),
\end{equation}
where $\Omega(\vX)$ is the number of microstates that are consistent with the macrostate $\vX$. Units such that $k_B = 1$ are used in order to simplify the notation; in this context, temperature has energy units. We use a tilde in order to identify statistical mechanics quantities. The thermodynamic entropy, homogeneous of order 1 in $N$, is $S(\vX)$. The difference is $\tilde{S}(\vX) - S(\vX) = O(\ln N)$.

First, we present the definition of Massieu functions in their thermodynamic (TH) and statistical mechanics (SM) versions for a specific situation: a particle subsystem of volume $V$ that is part of a larger complex; and the complement of the subsystem plays the role of a reservoir that imposes a temperature $T$ and a chemical potential $\mu$. Local fluctuations in $U$ and $N$ are present, and $V$ is a constant parameter by definition. We consider that the vector $\vX = (U,N)$ contains the extensive variables, but there can be other parameters, like $V$ in this case, that are fixed by external constraints. TH Massieu functions are obtained from successive applications of Legendre transforms, starting from the entropy. For example,
\begin{align}
	S(U,N) &  \nonumber \\
	S[\tfrac{1}{T},N] &= S(U,N) - \tfrac{1}{T} U \label{legendre} \\
	S[\tfrac{1}{T},\tfrac{\mu}{T}] &= S(U,N) - \tfrac{1}{T} U + \tfrac{\mu}{T} N. \nonumber
\end{align}
Different Massieu functions $S$ are identified by their variables between square brackets; the entropy, $S(U,N)$, with the variables in parenthesis, is the first Massieu function. We have that $S[\tfrac{1}{T},N] = -F/T$, where $F$ is the Helmholtz free energy, and $S[\tfrac{1}{T},\tfrac{\mu}{T}] = PV/T$, where $P$ is the pressure. A fundamental hypothesis of thermodynamics is that thermodynamic potentials, and TH Massieu functions, are homogeneous functions of order 1 in $N$; that is, they are extensive quantities. 

Now we turn to a statistical mechanics description. We have to specify how to calculate Massieu functions in this context. Instead of the Legendre transform, the mathematical transformation that generates SM Massieu functions starting from the entropy is the Laplace transform. Quantities with tilde, in the statistical mechanics context, contain nonextensive terms; we are interested in, for example, the difference $\tilde{S}[\tfrac{1}{T},\tfrac{\mu}{T}] - S[\tfrac{1}{T},\tfrac{\mu}{T}]$ at order $\ln N$. 

The TH Massieu functions $S[\tfrac{1}{T},N]$ and $S[\tfrac{1}{T},\tfrac{\mu}{T}]$, equal to $-F/T$ and $PV/T$, correspond to the logarithm of the canonical and grand canonical partition functions, $\mathcal{Z}(\tfrac{1}{T},N)$ and $\mathcal{Z}(\tfrac{1}{T},\tfrac{\mu}{T})$, respectively, as long as nonextensive terms are neglected in the thermodynamic limit. The SM Massieu functions are given by
\begin{align}
	\tilde{S}(\hat{U},\hat{N}) &= \ln \Omega(\hat{U},\hat{N})  \nonumber\\
	\tilde{S}[\tfrac{1}{T},\hat{N}] &= \ln \mathcal{Z}(\tfrac{1}{T},\hat{N}) \nonumber\\
	\tilde{S}[\tfrac{1}{T},\tfrac{\mu}{T}] &= \ln \mathcal{Z}(\tfrac{1}{T},\tfrac{\mu}{T}), \nonumber
\end{align}
where $\hat{U}$ and $\hat{N}$ are the energy and particle number for a given microstate $\omega$ in the grand canonical ensemble, where the reservoir fixes the values of $T$ and $\mu$. In the present notation, different partition functions are identified by their variables. Using the definitions of partition functions, we have
\begin{align}
	  e^{\tilde{S}(\hat{U},\hat{N})} &= \Omega(\hat{U},\hat{N}) \\
	e^{\tilde{S}[\frac{1}{T},\hat{N}]} &= \sum_{\hat{U}}  e^{\tilde{S}(\hat{U},\hat{N})} e^{-\frac{1}{T}\hat{U}} \label{tPsican} \\
	e^{\tilde{S}[\frac{1}{T},\frac{\mu}{T}]} &= \sum_{\hat{N}}\sum_{\hat{U}} e^{\tilde{S}(\hat{U},\hat{N})} e^{-\frac{1}{T}\hat{U} + \frac{\mu}{T}\hat{N}},
\end{align} 
where $\sum_{\hat{U}}$ and $\sum_{\hat{N}}$ represent the sum on all possible values of $\hat{U}$ and $\hat{N}$ (the sum is replaced by an integral for continuous variables). The mathematical transformation that generates partition functions (or SM Massieu functions) is the Laplace transform (see, for example, \cite[p.\ 247]{greiner}). The fundamental approximation that connects thermodynamics and statistical mechanics is to assume that there is one term that is overwhelmingly larger than the others in the sums (or integrals) above; we have this term when $\hat{U}$ and $\hat{N}$ take their thermodynamic average values $U$ and $N$, so we can write
\begin{align}
	e^{S(U,N)} &= \Omega(U,N) \\
	e^{S[\frac{1}{T},N]} &= e^{S(U,N)} e^{-\frac{1}{T} U} \\
	e^{S[\frac{1}{T},\frac{\mu}{T}]} &= e^{S(U,N)} e^{-\frac{1}{T} U + \frac{\mu}{T}N},
\end{align} 
and we recover the definitions of TH Massieu functions generated by Legendre transforms \eqref{legendre}. Since we have made an approximation, $\tilde{S}$ and $S$ are not exactly the same.

The probability of a microstate $\omega$ with energy $\hat{U}$ and particle number $\hat{N}$ is
\begin{equation}\label{PomegaUN}
	P(\omega) = \frac{\exp[- \tfrac{1}{T}\hat{U} + \tfrac{\mu}{T}\hat{N}]}{\mathcal{Z}(\tfrac{1}{T},\tfrac{\mu}{T})} = \exp(- \tilde{S}[\tfrac{1}{T},\tfrac{\mu}{T}] - \tfrac{1}{T}\hat{U} + \tfrac{\mu}{T}\hat{N}).
\end{equation}
In order to obtain the probability of the macrostate $\hat{\vX} = (\hat{U},\hat{N})$, we have to multiply by the corresponding number of compatible microstates $\Omega(\hat{U},\hat{N}) = \exp[\tilde{S}(\hat{U},\hat{N})]$; we have,
\begin{equation}\label{PUN}
    P(\hat{U},\hat{N}) = \exp[-\tilde{S}[\tfrac{1}{T},\tfrac{\mu}{T}] + \tilde{S}(\hat{U},\hat{N}) - \tfrac{1}{T}\hat{U} + \tfrac{\mu}{T}\hat{N}].
\end{equation}
Taking the sum of $P(\hat{U},\hat{N})$ over possible values of $\hat{U}$, we obtain the probability of $\hat{N}$:
\begin{equation}\label{PN}
	P(\hat{N}) = \exp[-\tilde{S}[\tfrac{1}{T},\tfrac{\mu}{T}] + \tilde{S}[\tfrac{1}{T},\hat{N}] + \tfrac{\mu}{T}\hat{N}],
\end{equation}
where Eq.\ \eqref{tPsican} was used.

In general, for a given microstate $\omega$ the extensive variables take the values $\hat{\vX}=(\hat{X}_1,\dots,\hat{X}_r)$; the list does not include extensive parameters fixed by external constraints; variables without a hat, $X_i$, are the thermodynamic averages. The probability of the microstate $\omega$ is given by (see, for example, \cite[Ch.\ 19]{callen})
\begin{equation}\label{Pomega}
	P(\omega) = \exp[-\tilde{S}[\vY] - \vY\cdot\hat{\vX}],
\end{equation}
where $\vY = (Y_1,\dots,Y_r)$ are entropic intensive parameters defined by $Y_i = \frac{\partial S(\vX)}{\partial X_i}$ and set by the reservoir. Multiplying by the number of microstates with $\hat{\vX}$, given by $\Omega(\hat{\vX}) = \exp[\tilde{S}(\hat{\vX})]$, we have the probability of $\hat{\vX}$:
\begin{equation}\label{PX}
	P(\hat{\vX}) = \exp[-\tilde{S}[\vY] + \tilde{S}(\hat{\vX}) - \vY\cdot\hat{\vX}].
\end{equation}

If we have $s$ extensive variables and $r-s$ intensive parameters, that we can write as $\vX^s = (X_1,\dots,X_s)$ and $\vY^{r-s}=(Y_{s+1},\dots,Y_r)$, then the SM Massieu function is the logarithm of the partition function in the corresponding ensemble:
\begin{equation}\label{MassieuST}
	\tilde{S}[\vX^s,\vY^{r-s}] = \ln \mathcal{Z}(\vX^s,\vY^{r-s}),
\end{equation}
where we have used the average values of the extensive variables (without a hat), but the same expression holds for $\hat{\vX}^s$ for a given microstate. Also, the SM Massieu function that depends on $s$ extensive variables and $r-s$ intensive parameters is given by $r-s$ Laplace transforms of $\tilde{S}(\hat{\vX})$:
\begin{align}\label{Psigen}
	e^{\tilde{S}[\hat{\vX}^s, \vY^{r-s}]} &= \\ \nonumber \sum_{\hat{X}_{s+1}} \dots \sum_{\hat{X}_r}& e^{\tilde{S}(\hat{\vX})} e^{-Y_{s+1}\hat{X}_{s+1} \dots - Y_r\hat{X}_r}.
\end{align}
The probability of a subset of extensive parameters, $\hat{X}_1,\dots,\hat{X}_s$, is given by
\begin{align}\label{PXs}
	P(\hat{\vX}^s) &= \sum_{\hat{X}_{s+1}} \dots \sum_{\hat{X}_r} P(\hat{\vX}) \nonumber\\ 
	&= e^{-\tilde{S}[\vY] + \tilde{S}[\hat{\vX}^s, \vY^{r-s}] - Y_1 \hat{X}_1 \dots - Y_s \hat{X}_s},
\end{align}
where Eqs.\ \eqref{PX} and \eqref{Psigen} were used.

\section{nonextensive term and thermodynamic fluctuations}
\label{s.nonext}

In this section we demonstrate that the difference between thermodynamic and statistical mechanics Massieu functions is given by thermodynamic fluctuations.

\subsection{$\hat{U},\hat{N}, V$ system}

Let us first consider a simple specific case: We want to calculate the difference $\tilde{S}[\tfrac{1}{T},N] - S[\tfrac{1}{T},N]$ for the system in contact to energy and particle reservoirs. That is, the difference between the statistical mechanics free energy, obtained from the partition function, and the thermodynamic free energy, since $\tilde{S}[\tfrac{1}{T},N] = \ln\mathcal{Z}(\frac{1}{T},N) = -\tilde{F}/T$ and $S[\tfrac{1}{T},N] = -F/T$. It can be shown that both quantities are related through their derivatives. By definition, we have
\begin{equation}\label{dpsit}
	\frac{\partial S[\tfrac{1}{T},N]}{\partial N} = - \frac{\mu}{T}.
\end{equation}
Even for small systems, it is usually assumed that the particle number $N$ is large enough so that thermodynamic potentials can be taken as continuous functions of $N$ \cite[p.\ 6]{hill2}. On the other hand, we can calculate the average of the derivative of $\tilde{S}[\tfrac{1}{T},\hat{N}]$. Taking the logarithm of Eq.\ \eqref{PN} we have,
\begin{equation}\label{Psi3}
	\tilde{S}[\tfrac{1}{T},\hat{N}] = \ln P(\hat{N}) + \tilde{S}[\tfrac{1}{T},\tfrac{\mu}{T}] - \tfrac{\mu}{T}\hat{N}.
\end{equation}
The average of its derivative is
\begin{align}\label{avder}
	\left\langle \frac{\partial \tilde{S}[\tfrac{1}{T},\hat{N}]}{\partial \hat{N}}\right\rangle &= \left\langle \frac{1}{P(\hat{N})}\frac{\partial P(\hat{N})}{\partial \hat{N}} - \frac{\mu}{T} \right\rangle \nonumber\\
	&= - \frac{\mu}{T} + \sum_{\hat{N}}\frac{\partial P(\hat{N})}{\partial \hat{N}}.  
\end{align}
The sum can be transformed into an integral writing the probability in terms of the probability density, $P(\hat{N}) = D(\hat{N})\,d\hat{N}$:
\begin{align}\label{avder2}
	\left\langle \frac{\partial \tilde{S}[\tfrac{1}{T},\hat{N}]}{\partial \hat{N}}\right\rangle &= - \frac{\mu}{T} + \int_0^{N_{\rm max}} d\hat{N}\;\frac{\partial D(\hat{N})}{\partial \hat{N}} \nonumber\\
	&= - \frac{\mu}{T} + \cancel{D(N_{\rm max})} - \cancel{D(0)}.
\end{align}
Terms $D(N_{\rm max})$ and $D(0)$ were canceled since the probability distribution vanishes at the extreme values of the variable. Then, from \eqref{dpsit} and \eqref{avder2} we have that 
\begin{equation}\label{der1}
	\left\langle \frac{\partial \tilde{S}[\tfrac{1}{T},\hat{N}]}{\partial \hat{N}}\right\rangle = \frac{\partial S[\tfrac{1}{T},N]}{\partial N}.
\end{equation}
The average can be written in terms of mean-squared fluctuations if we expand $\frac{\partial \tilde{S}[\frac{1}{T},\hat{N}]}{\partial \hat{N}}$ in a Taylor series around the mean value $N$:
\begin{equation}\label{der2}
	\left\langle \frac{\partial \tilde{S}[\tfrac{1}{T},\hat{N}]}{\partial \hat{N}}\right\rangle = \frac{\partial \tilde{S}[\tfrac{1}{T},N]}{\partial N} + \frac{1}{2}\frac{\partial^3 S[\tfrac{1}{T},N]}{\partial N^3} \langle (\Delta \hat{N})^2\rangle,
\end{equation}
where $\Delta \hat{N} = \hat{N} - N$, and the terms $O(N^{-2})$ are neglected [the mean squared or higher-order fluctuations are $O(N)$, and the $i$th derivative is $O(N^{-i+1})$]; the third derivative of $\tilde{S}$ was replaced by the third derivative of $S$ (without tilde) since the difference contributes with a term $O(N^{-2})$.  Combining the last two equations,
\begin{equation}\label{dif1}
	\frac{\partial}{\partial N}(\tilde{S}[\tfrac{1}{T},N] - S[\tfrac{1}{T},N]) = -\frac{1}{2}\frac{\partial^3 S[\tfrac{1}{T},N]}{\partial N^3} \langle (\Delta \hat{N})^2\rangle.
\end{equation}
The fluctuations are (see, for example, \cite{mishin} or  Sec.\ 19.3 in \cite{callen})
\begin{equation}\label{fluctN}
	\langle (\Delta \hat{N})^2\rangle = - \frac{\partial N}{\partial (-\tfrac{\mu}{T})} = - \frac{1}{\frac{\partial (-\tfrac{\mu}{T})}{\partial N}} = - \frac{1}{\frac{\partial^2 S[\frac{1}{T},N]}{\partial N^2}},  
\end{equation}
then,
\begin{equation}\label{der3ra}
	\frac{\partial^3 S[\tfrac{1}{T},N]}{\partial N^3} = \frac{1}{\langle (\Delta \hat{N})^2\rangle^2} \frac{\partial \langle (\Delta \hat{N})^2\rangle}{\partial N}.
\end{equation}
Replacing \eqref{der3ra} in \eqref{dif1}, we obtain
\begin{equation}\label{dif2}
	\frac{\partial \Delta\tilde{S}[\tfrac{1}{T},N]}{\partial N} = -\frac{1}{2}\frac{\partial}{\partial N} \ln\langle (\Delta \hat{N})^2\rangle,
\end{equation}
where $\Delta \tilde{S}[\tfrac{1}{T},N] = \tilde{S}[\tfrac{1}{T},N] - S[\tfrac{1}{T},N]$; and, integrating in $N$,
\begin{equation}\label{dif3}
	\Delta \tilde{S}[\tfrac{1}{T},N] = -\frac{1}{2}\ln\langle (\Delta \hat{N})^2\rangle + c,
\end{equation}
where $c$ is an integration constant [$c$ may be neglected respect to the logarithm of fluctuations $O(\ln N)$]. Let us notice that this expression was obtained by Lebowitz and Percus in terms of the free energy for a fluid particle system; see Eq.\ (3.11) in Ref.\ \cite{lebowitz-percus}. The present development based on Massieu functions allows the derivation of more general expressions, as shown in the next section.

If interactions among components of the system can be neglected, the TH and SM Massieu functions take their ideal values, $S_\text{id}[\tfrac{1}{T},N]$ and $\tilde{S}_\text{id}[\tfrac{1}{T},N]$. Assuming the ideal situation, Eq.\ \eqref{dif3} is
\begin{equation}\label{dif3id}
	\Delta \tilde{S}_\text{id}[\tfrac{1}{T},N] = -\frac{1}{2}\ln\langle (\Delta \hat{N})^2\rangle_\text{id} + c.
\end{equation}

The excess Massieu function (TH or SM) is defined as the difference $S_\text{ex} = S - S_\text{id}$ (with or without a tilde). Doing subtraction between Eqs.\ \eqref{dif3} and \eqref{dif3id}, we obtain,
\begin{equation}\label{difex}
	\Delta\tilde{S}_\text{ex}[\tfrac{1}{T},N] = -\frac{1}{2}\ln\frac{\langle (\Delta \hat{N})^2\rangle}{\langle (\Delta \hat{N})^2\rangle_\text{id}}.
\end{equation}
Then, the nonextensive term of the excess part does not behave as $\ln N$ but, instead, it is $O(N^0)$.

\subsection{General derivations}

Let us consider the system with $s$ extensive and $r-s$ intensive variables given by $\hat{\vX}^s$ and $\vY^{r-s}$. From Eq.\ \eqref{PXs}, we have the SM Massieu function
\begin{equation}\label{Psis}
	\tilde{S}[\hat{\vX}^s, \vY^{r-s}] = \ln P(\hat{\vX}^s) + \tilde{S}[\vY] + Y_1 \hat{X}_1 + \dots + Y_s \hat{X}_s.
\end{equation}
The average of its derivative respect to $\hat{X}_i$, with $i=1,\dots, s$, is
\begin{align}\label{aPsis}
		\left\langle \frac{\partial \tilde{S}[\hat{\vX}^s, \vY^{r-s}]}{\partial \hat{X}_i}\right\rangle &= Y_i + \left\langle \frac{1}{P(\hat{\vX}^s)}\frac{\partial P(\hat{\vX}^s)}{\partial \hat{X}_i} \right\rangle \nonumber\\
	&= Y_i + \sum_{\hat{X}_1}\dots \sum_{\hat{X}_s}\frac{\partial P(\hat{\vX}^s)}{\partial \hat{X}_i} \nonumber\\
	&= Y_i + \sum_{\hat{X}_i}\frac{\partial P(\hat{X}_i)}{\partial \hat{X}_i} \nonumber\\
	&= Y_i,  \qquad i=1,\dots, s
\end{align}
where, in the last line, the procedure used in Eq.\ \eqref{avder2} was followed: the sum is transformed into an integral and it is considered that the probability density vanishes at the extreme values of $\hat{X}_i$. Then,
\begin{equation}\label{aPsis2}
	\left\langle \frac{\partial \tilde{S}[\hat{\vX}^s, \vY^{r-s}]}{\partial \hat{X}_i}\right\rangle = \frac{\partial S[\vX^s, \vY^{r-s}]}{\partial X_i}.
\end{equation}
A Taylor expansion around $\vX^s$ is used to evaluate the average:
\begin{align}\label{avs}
	\left\langle \frac{\partial \tilde{S}[\hat{\vX}^s, \vY^{r-s}]}{\partial \hat{X}_i}\right\rangle &= \frac{\partial \tilde{S}[\vX^s, \vY^{r-s}]}{\partial X_i} \nonumber\\ 
	& +\frac{1}{2} \sum_{j,k} \frac{\partial^3 S[\vX^s, \vY^{r-s}]}{\partial X_i \partial X_j \partial X_k} \langle \Delta \hat{X}_j \Delta \hat{X}_k\rangle,
\end{align}
with $\Delta \hat{X}_j = \hat{X}_j - X_j$ and terms $O(N^{-2})$ are neglected (including the difference between $S$ and $\tilde{S}$ in the third derivative). Therefore,
\begin{align}\label{difPsis}
	\frac{\partial}{\partial X_i}&(\tilde{S}[\vX^s, \vY^{r-s}] - S[\vX^s, \vY^{r-s}]) \nonumber\\
	&= -\frac{1}{2} \sum_{j,k} \frac{\partial^3 S[\vX^s, \vY^{r-s}]}{\partial X_i \partial X_j \partial X_k} \langle \Delta \hat{X}_j \Delta \hat{X}_k\rangle \nonumber\\
    &= -\frac{1}{2} \sum_{j,k} \frac{\partial^2 Y_j}{\partial X_i\partial X_k} \langle \Delta \hat{X}_j \Delta \hat{X}_k\rangle.
\end{align}

Let us call $C$ the (symmetric) fluctuation matrix. The component $C_{jk}$ is given by (see \cite{callen})
\begin{equation}\label{fluct2}
	C_{jk} = \langle \Delta \hat{X}_j \Delta \hat{X}_k\rangle = -\frac{\partial X_j}{\partial Y_k}.
\end{equation}
The inverse of the fluctuation matrix has components $(C^{-1})_{jk} = -\frac{\partial Y_j}{\partial X_k}$.
Now, Eq.\ \eqref{difPsis} can be written in terms of the components of $C$ and its inverse, $C^{-1}$ (also symmetric): 
\begin{align}\label{difPsis2}
    \frac{\partial \Delta\tilde{S}[\vX^s, \vY^{r-s}]}{\partial X_i} &= \frac{1}{2} \sum_{j,k} \frac{\partial (C^{-1})_{jk}}{\partial X_i} C_{jk} \nonumber\\
    &= \frac{1}{2} {\rm tr}\left( \frac{\partial C^{-1}}{\partial X_i} \cdot C \right), \qquad i=1,\dots, s
\end{align}
where we define $\Delta \tilde{S}[\vX^s, \vY^{r-s}] \equiv \tilde{S}[\vX^s, \vY^{r-s}] - S[\vX^s, \vY^{r-s}]$. It can be shown (see Sec.\ 15.8 in \cite{harville}) that the solution of this equation system is
\begin{equation}\label{difPsis3}
	\Delta \tilde{S}[\vX^s, \vY^{r-s}] = -\frac{1}{2}  \ln (\det C) + c,
\end{equation}
where $c$ is an integration constant ($c$ should satisfy $\frac{\partial c}{\partial X_1} = \dots = \frac{\partial c}{\partial X_s} = 0$). For the ideal case, Eq.\ \eqref{difPsis3} becomes
\begin{equation}\label{difPsisid}
    \Delta \tilde{S}_\text{id}[\vX^s, \vY^{r-s}] = -\frac{1}{2}  \ln (\det C_\text{id}) + c.
\end{equation}
Subtracting the last two equations, we obtain the nonextensive term of the excess Massieu function:
\begin{align}
    \Delta \tilde{S}_\text{ex}[\vX^s, \vY^{r-s}] &= \tilde{S}_\text{ex}[\vX^s, \vY^{r-s}] - S_\text{ex}[\vX^s, \vY^{r-s}]  \nonumber\\
	&= -\frac{1}{2}  \ln \left(\frac{\det C}{\det C_\text{id}}\right). \label{difPsiex}
\end{align}
If the only extensive variable that fluctuates is the number of particles, $\hat{N}$, then this equation reduces to Eq.\ \eqref{difex}.

Using the result of Eq.\ \eqref{difPsis3}, and the normalization of the probability distribution of fluctuations, it can be shown that the difference $\Delta\tilde{S}[\vY]$ for the last Legendre transform is a constant; see Appendix A. It is known that this constant is zero for some solvable systems, meaning that $\tilde{S}[\vY]$ is exactly extensive; this is the case for the one-dimensional spin lattice, as is mentioned in Sec.\ \ref{s.spinnonideal} below, and for the grand canonical ensemble of an ideal gas (See Sec.\ 8.2 in \cite{swendsen2}, where the Massieu function that depends on $\vY=(\frac{1}{T},\frac{\mu}{T})$ is called the grand canonical entropy.)

\section{Particle system}
\label{s.particles}

In this section we verify previous results for a system of particles with and without interactions.

\subsection{Ideal case}

For $N$ noninteracting particles in a volume $V$ at temperature $T$, the ideal canonical partition function is $\mathcal{Z}_\text{id} = (V/\lambda^3)^N/N!$, where $\lambda$ is the thermal de Broglie wavelength. The SM Massieu function is
\begin{equation}\label{Psiidealsm}
	\tilde{S}_\text{id}[\tfrac{1}{T},N] = \ln \mathcal{Z}_\text{id} = N \ln\left(\frac{V}{\lambda^3 N}\right)  + N - \frac{1}{2}\ln (2\pi N)
\end{equation}
where Stirling's approximation was used. In the context of thermodynamics, the TH Massieu function is an extensive quantity:
\begin{equation}\label{Psiidealth}
	S_\text{id}[\tfrac{1}{T},N] = N \ln\left(\frac{V}{\lambda^3 N}\right) + N
\end{equation}
The difference $\tilde{S}_\text{id}[\tfrac{1}{T},N] - S_\text{id}[\tfrac{1}{T},N]$ is
\begin{equation}\label{difid}
	\Delta\tilde{S}_\text{id}[\tfrac{1}{T},N] = -\frac{1}{2}\ln N -\frac{1}{2}\ln (2\pi).
\end{equation}
This result satisfies Eq.\ \eqref{dif3id} [with $c = -\frac{1}{2}\ln (2\pi)$], since the mean-squared fluctuations of particle number in the ideal system are $\langle (\Delta \hat{N})^2\rangle = N$.

\subsection{Non ideal case}

In this section we numerically check the validity of the results for the nonextensive term of the excess Massieu function when interactions between particles are present.

As before, we analyze the system of particles described by $\hat{U}$, $\hat{N}$, and $V$, where energy and particle number fluctuate and volume is constant. We are interested in the nonextensive term of the Massieu function that corresponds to the excess free energy: $\tilde{S}_\text{ex}[\tfrac{1}{T},N] = - \tilde{F}_\text{ex}/T$. This quantity is related to the particle fluctuations, as can be seen in Eq.\ \eqref{difex}. The mean-squared fluctuations are [see Eq.\ \eqref{fluctN}] 
\begin{equation}\label{DeltaN}
	\langle (\Delta \hat{N})^2\rangle = \frac{N}{\Gamma},
\end{equation}
where $\Gamma = \frac{N}{T}\frac{\partial \mu}{\partial N}$ is the thermodynamic factor; for the ideal case, $\Gamma=1$. Then, using \eqref{difex},
\begin{equation}\label{difex2}
	\Delta\tilde{S}_\text{ex}[\tfrac{1}{T},N] = \frac{1}{2} \ln \Gamma,
\end{equation}
or, in terms of the excess free energy,
\begin{equation}\label{difex3}
	\frac{\tilde{F}_\text{ex}}{T} - \frac{F_\text{ex}}{T} = -\frac{1}{2} \ln \Gamma.
\end{equation}
The excess chemical potential is $\mu_\text{ex} = \frac{\partial F_\text{ex}}{\partial N}$, and the one including nonextensive terms is $\tilde{\mu}_\text{ex} = \frac{\partial \tilde{F}_\text{ex}}{\partial N}$. Then, taking the derivative of the previous equation, we obtain
\begin{equation}\label{muex}
	\tilde{\mu}_\text{ex} - \mu_\text{ex} = -\frac{T}{2} \frac{\Gamma'}{\Gamma},
\end{equation}
with $\Gamma'=\frac{\partial\Gamma}{\partial N}$ [see Eq.\ (B10) in Ref.\ \cite{dimuro}]. When the excess chemical potential is obtained numerically (using, for example, the Widom insertion method), an additional term $O(N^{-1})$ has to be taken into account: $\tilde{\mu}_\text{ex}^\text{num} = \tilde{F}_\text{ex}(N+1) - \tilde{F}_\text{ex}(N) = \tilde{\mu}_\text{ex} + \tilde{\mu}_\text{ex}'/2 + O(N^{-2})$, with  $\tilde{\mu}_\text{ex}' = \frac{\partial \tilde{\mu}_\text{ex}}{\partial N}$. Then, Eq.\ \eqref{muex} becomes
\begin{equation}\label{muex2}
	\tilde{\mu}_\text{ex}^\text{num} - \mu_\text{ex} = -\frac{T}{2} \frac{\Gamma'}{\Gamma} + \frac{1}{2} \mu_\text{ex}',
\end{equation}
where we used $\tilde{\mu}_\text{ex}' = \mu_\text{ex}' + O(N^{-2})$, and terms $O(N^{-2})$ are neglected.
As mentioned in the Introduction, Siepmann \textit{et al.}\ \cite{siepmann} analyzed finite-size corrections to the excess chemical potential in a particle system with fixed volume and periodic boundary conditions. They found that 
\begin{equation}\label{siep}
    \tilde{\mu}_\text{ex}^\text{num} - \mu_\text{ex} = \frac{1}{2N}\frac{\partial P}{\partial \rho}\left[ 1 - T\frac{\partial \rho}{\partial P} - \rho T \frac{\partial^2 P}{\partial \rho^2} \left( \frac{\partial \rho}{\partial P} \right)^2 \right],
\end{equation}
where $P$ is the pressure and $\rho=N/V$ is the density. It can be seen that this equation is equivalent to \eqref{muex2}. Using the thermodynamic relationship $N\mu = F + PV$ and the definition of the thermodynamic factor, we have $\Gamma = \frac{N}{T}\frac{\partial \mu}{\partial N} = \frac{1}{T} \frac{\partial (PV)}{\partial N} = \frac{1}{T} \frac{\partial P}{\partial \rho}$. Rewriting Eq.\ \eqref{siep} in terms of $\Gamma$, and knowing that $\Gamma = 1 + \frac{N}{T}\mu_\text{ex}'$, after some algebra we recover Eq.\ \eqref{muex2}.

Next, we contrast the analytical expression against numerical results for hard disks and spheres. 
Let us consider first hard disk interaction. Solana have proposed a simple equation of state (EOS) for a system of hard disks \cite{Solana},
\begin{equation}\label{e.Solana}
Z=\frac{1+\eta^2/8-\eta^4/10}{(1-\eta)^2},
\end{equation}
where $Z$ is the compressibility factor and $\eta = \frac{N}{V} \frac{\pi}{4}\sigma^2$ is the packing fraction, with $\sigma$ the particle diameter. The compressibility factor is defined as $Z = \frac{PV}{N T}$, where, as mentioned before, temperature is in energy units. Thermodynamic properties are obtained from the EOS (see, for example, Sec.\ IV in \cite{marchioni}). In particular, it can be shown  that the thermodynamic factor and the excess free energy are
\begin{align}
\Gamma &= \frac{40+40\eta+15\eta^2-5\eta^3-20\eta^4+12\eta^5}{40(1-\eta)^3} , \\
\frac{F_\text{ex}}{TN} &=1.025\Big(\frac{1}{1-\eta}-1\Big)-0.2\eta-0.05\eta^2 \nonumber\\
& \ \ \ -1.175\ln(1-\eta).
\end{align}
On the other hand, $\tilde{F}_\text{ex}/T$ is obtained from the partition function:
\begin{equation}\label{FexZ}
	\frac{\tilde{F}_\text{ex}}{T} = -\ln\frac{\mathcal{Z}}{\mathcal{Z}_\text{id}} = -\ln \langle e^{-\phi/T}\rangle_0,
\end{equation}
where $\phi$ is the interaction energy that depends on all particle positions, and $\langle\rangle_0$ denotes the average over non-interacting particle configurations (see, for example, Sec.\ 5.1 in \cite{kardar}).

\begin{figure}
    \centering
    \includegraphics[width=\linewidth]{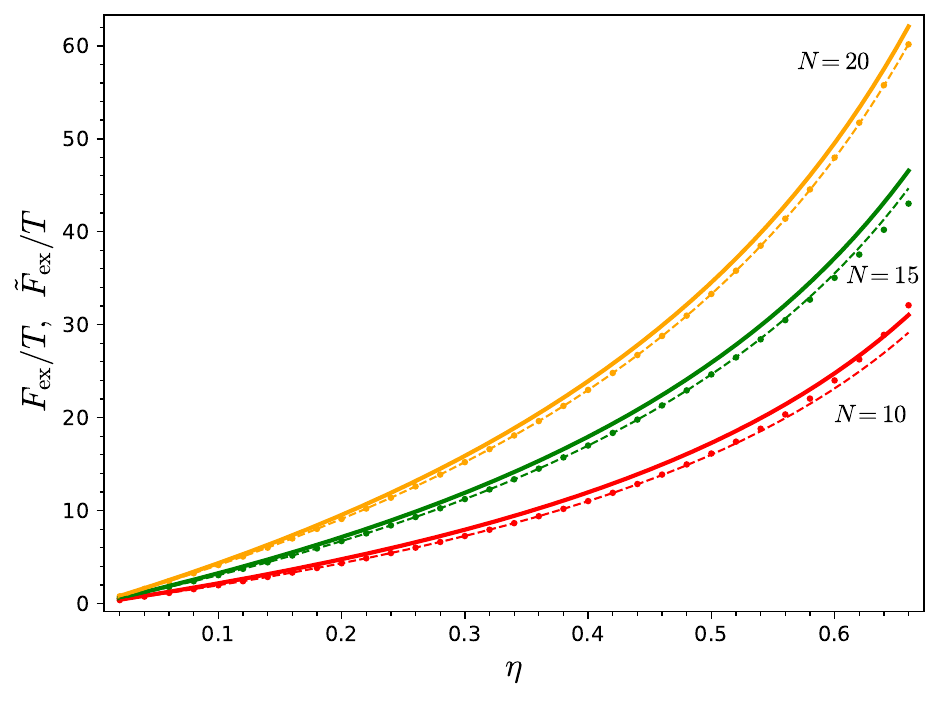}
    \caption{Nonextensive term $\tilde{F}_{\rm ex}/T$ (dots) and $F_{\rm ex}/T$ (solid line)   as a function of the packing fraction $\eta$ for a system of hard disks. The dashed lines represent $F_{\rm ex}/T-1/2\log{\Gamma}$ [see Eq.\ (\ref{difex3})]. $N$ is the number of particles. }
    \label{fig:HD_F}
\end{figure}

\begin{figure}
    \centering
    \includegraphics[width=\linewidth]{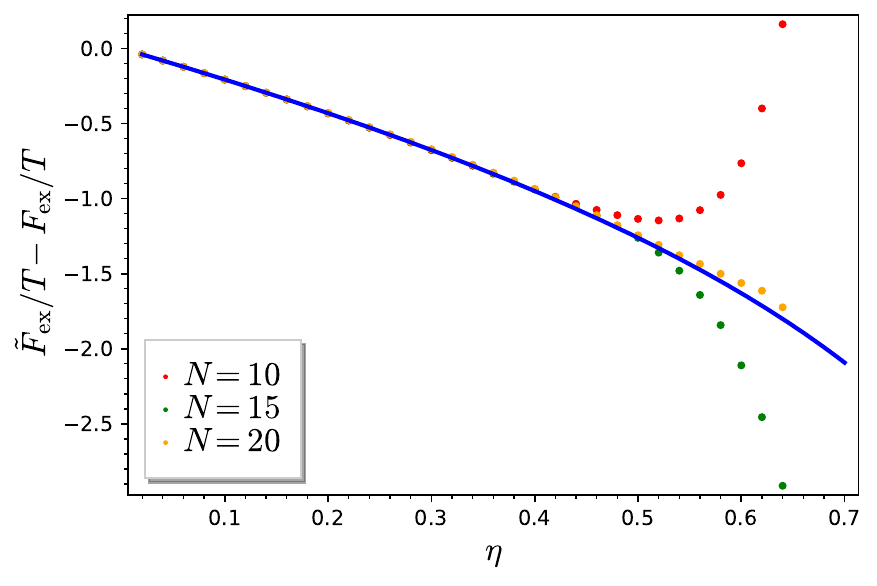}
    \caption{Nonextensive term $\tilde{F}_{\rm ex}/T-F_{\rm ex}/T$  against the packing fraction $\eta$ for a system of hard disks. The curve corresponds to the right hand side of Eq.\ (\ref{difex3}). Symbols represent numerical results; $N$ is the number of particles. The statistical error does not exceed $0.01 \%$.}
    \label{fig:HD}
\end{figure}

\begin{figure}
    \centering
    \includegraphics[width=\linewidth]{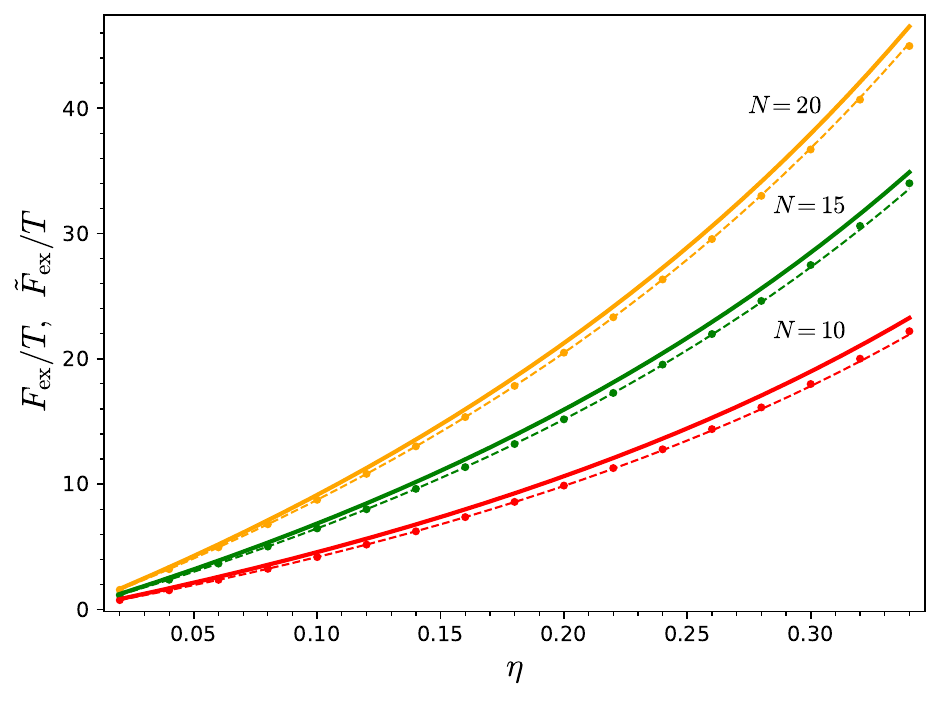}
    \caption{Nonextensive term $\tilde{F}_{\rm ex}/T$ (dots) and $F_{\rm ex}/T$ (solid line) against the packing fraction $\eta$ for a system of hard spheres. The dashed lines represent $F_{\rm ex}/T-1/2\log{\Gamma}$ [see Eq.\ (\ref{difex3})]. $N$ is the number of particles.}
    \label{fig:HS_F}
\end{figure}

\begin{figure}
    \centering
    \includegraphics[width=\linewidth]{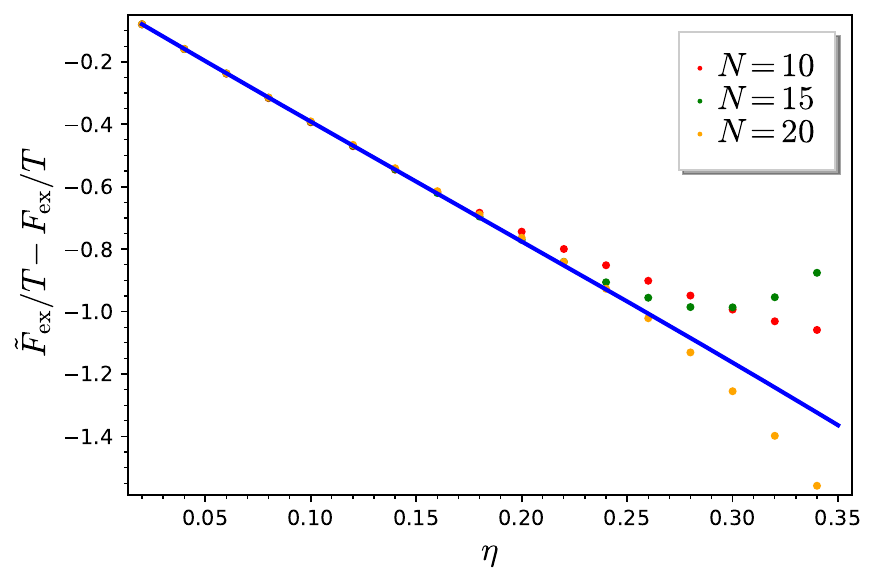}
    \caption{Nonextensive term $\tilde{F}_{\rm ex}/T-F_{\rm ex}/T$  against the packing fraction $\eta$ for a system of hard spheres. The curve corresponds to Eq.\ (\ref{difex3}). Dots represent numerical results; $N$ is the number of particles. The statistical error is smaller than $0.01 \%$.}
    \label{fig:HS}
\end{figure}

This average can be numerically obtained via Monte Carlo simulations; see Appendix B. 

In Fig.\ \ref{fig:HD_F} we compare the thermodynamic and statistical mechanics excess free energy, $\tilde{F}_\text{ex}$ and $F_\text{ex}$, respectively, for $N=10,15$, and $20$ particles The dashed curve is the correction to $F_\text{ex}$, given by Eq.\ (\ref{difex3}). In Fig.\ \ref{fig:HD} we show the correction given by Eq.\ (\ref{difex3}). We can observe in Fig.\ \ref{fig:HD} that the agreement with the theoretical prediction of Eq.\ (\ref{difex3}) improves as the number of particles increases. For instance, for $\eta=0.4$ the deviations are around $8\%$, $5\%$, and $4\%$ for $N=10$, $15$, and $20$, respectively. 

Next we analyze hard sphere interaction. The EOS obtained by Carnahan and Starling \cite{carnahan} is widely used for hard spheres in the fluid phase. It is given by
\begin{equation}\label{e.eos}
	Z = \frac{1 + \eta + \eta^2 - \eta^3}{(1-\eta)^3},
\end{equation}
where $\eta = \frac{N}{V} \frac{\pi}{6}\sigma^3$ is the packing fraction, with $\sigma$ the particle diameter. In this case the thermodynamic factor and the excess free energy are
\begin{align}
	\Gamma &= \frac{\eta^{4} - 4 \, \eta^{3} + 4 \, \eta^{2} + 4 \, \eta + 1}{\eta^{4} - 4 \, \eta^{3} + 6 \, \eta^{2} - 4 \, \eta + 1}, \\
	\frac{F_\text{ex}}{T} &= -N\frac{3 \, \eta^{2} - 4 \, \eta}{\eta^{2} - 2 \, \eta + 1}.
\end{align}

In Fig.\ \ref{fig:HS_F} we show numerical results for the SM and TH free energy, for $N=10,15$, and $20$ particles. In Fig.\ \ref{fig:HS} we can see in more detail the correction of Eq.\ (\ref{difex3}).  As we observed before for hard disks, the agreement with the theoretical line also improves for larger system size. Here, for $\eta=0.2$ the deviations are around $7\%$, $5\%$, and $4\%$ for $N=10$, $15$, and $20$, respectively. Other EOS were also considered \cite{boublik2,kolafa2}, but no significant difference was found.

\section{Spin lattice}
\label{s.spin}

Let us consider a $d$-dimensional lattice with $N$ spins, where each spin has two possible states: up or down. There are $\hat{N}_+$ and $\hat{N}_-=N - \hat{N}_+$ spins in the up and down states. The total number of spins, $N$, is constant and plays the same role as the volume in the previous example. Each spin, $s_i$, takes the value 1 or $-1$. The microscopic state is given by $\vec{s} = (s_1,\dots,s_N)$ and its energy is 
\begin{equation}
    \hat{H} = \hat{E} - h \hat{M},
\end{equation}
where $\hat{E} = -\epsilon \sum_{\langle i,j\rangle} s_i s_j$ is the interaction energy, $\sum_{\langle i,j\rangle}$ is the sum over nearest neighbor pairs, and $\hat{M} = \sum_i s_i$ is the magnetization; $\epsilon$ is the coupling energy and $h$ represents an external magnetic field. 

The set of SM Massieu functions (for constant spin number $N$) is
\begin{align}
	e^{\tilde{S}(\hat{M},\hat{E})} &= \Omega(\hat{M},\hat{E}),\\
	e^{\tilde{S}[\hat{M},\frac{1}{T}]} &= \sum_{\hat{E}} e^{\tilde{S}(\hat{M},\hat{E})} e^{-\frac{1}{T}\hat{E}} \label{PMT0},\\
	e^{\tilde{S}[\frac{h}{T},\frac{1}{T}]} &= \sum_{\hat{M}} e^{\tilde{S}[\hat{M},\frac{1}{T}]} e^{\frac{h}{T}\hat{M}}. \label{PhT0}
\end{align} 

\subsection{Ideal case}

In the ideal case, we have $\epsilon=0$. The macroscopic state is represented by the magnetization $\hat{M} = \sum_i s_i = \hat{N}_+ - \hat{N}_-$ and the total spin number $N$; the number of spins up and down are $\hat{N}_+ = (N + \hat{M})/2$ and $\hat{N}_- = (N - \hat{M})/2$.

Equation\ \eqref{PhT0} reduces to 
\begin{equation}\label{PsimZ}
	e^{\tilde{S}_\text{id}[\frac{h}{T}]} = \sum_{\hat{M}} e^{\tilde{S}_\text{id}(\hat{M})} e^{\frac{h}{T} \hat{M}},
\end{equation}
where
\begin{equation}\label{Psim}
	\tilde{S}_\text{id}(\hat{M})  = \ln \Omega(\hat{M}) = \ln \frac{N!}{\hat{N}_+!\hat{N}_-!},
\end{equation}
with $\Omega(\hat{M})$ the number of microstates with magnetization $\hat{M}$.
Let us consider $\tilde{S}_\text{id}(M)$, evaluated at the thermodynamic average $M$, and apply Stirling's approximation (up to order $\ln N$):
\begin{equation}\label{Psim2}
	\tilde{S}_\text{id}(M) = \underbrace{N_+ \ln \frac{N}{N_+} + N_- \ln \frac{N}{N_-}}_{O(N)} - \underbrace{\frac{1}{2} \ln \frac{N_+ N_-}{N}}_{O(\ln N)}.
\end{equation}
The extensive part, order $N$, is the corresponding TH Massieu function:
\begin{equation}\label{Psim2t}
	S_\text{id}(M) = N_+ \ln \frac{N}{N_+} + N_- \ln \frac{N}{N_-}.
\end{equation}
So, the difference is
\begin{equation}\label{difM}
	\Delta\tilde{S}_\text{id}(M) = -\frac{1}{2} \ln \frac{N_+ N_-}{N}.
\end{equation}
According to Eq.\ \eqref{difPsisid}, this difference should be given by the mean squared fluctuations of $\hat{M}$,
\begin{equation}\label{difM2}
    \Delta\tilde{S}_\text{id}(M) = - \frac{1}{2} \ln \langle (\Delta\hat{M})^2\rangle_\text{id},
\end{equation}
where the terms $O(N^0)$ are neglected (as in Stirling's approximation above). The fluctuations are
\begin{equation}\label{fluctM}
	\langle (\Delta\hat{M})^2\rangle_\text{id} = \frac{\partial M}{\partial \frac{h}{T}} = \frac{-1}{\frac{\partial^2 S_\text{id}(M)}{\partial M^2}} = 4\frac{N_+ N_-}{N},
\end{equation}
where we have used that $\frac{h}{T}= -\frac{\partial S_\text{id}(M)}{\partial M}$ (a relationship that can be obtained from the Legendre transform $S_\text{id}[\frac{h}{T}] = S_\text{id}(M) + \frac{h}{T}M$). 
Therefore, comparing Eqs.\ \eqref{difM} and \eqref{difM2}, it can be seen that the theory correctly predicts the nonextensive term of the Massieu function at order $\ln N$.

\subsection{Nonideal case}
\label{s.spinnonideal}

By applying the thermodynamic limit approximation to Eq.\ \eqref{PhT0}, we have the Legendre transform
\begin{equation}\label{PMJ}
    S[M,\tfrac{1}{T}] = S[\tfrac{h}{T},\tfrac{1}{T}] - \tfrac{h}{T}M.
\end{equation}
$S[M,\tfrac{1}{T}]$ and $S[\tfrac{h}{T},\tfrac{1}{T}]$ are associated to the Helmholtz and Gibbs free energies of the spin lattice, respectively. Let us consider the exact solution of the Ising model of $N$ spins in one dimension with periodic boundary conditions; $\ln \mathcal{Z}(\tfrac{h}{T},\tfrac{1}{T})$ is extensive, so $S[\tfrac{h}{T},\tfrac{1}{T}]=\tilde{S}[\tfrac{h}{T},\tfrac{1}{T}]$ (see, for example, Sec.\ 5.6.1 in  \cite{reichl} or Sec.\ 6.5.2 in \cite{schwabl}). The result is
\begin{align}
    S(&\tfrac{h}{T},\tfrac{1}{T}) =  N \frac{\epsilon}{T} \nonumber\\
    &+ N\ln\left[\cosh \left(\tfrac{h}{T}\right) + \sqrt{\cosh^2 \left(\tfrac{h}{T}\right) - 2 e^{-2\epsilon/T} \sinh\left( \tfrac{2\epsilon}{T} \right)}\right]. \label{PhJ}
\end{align}
From $M = \frac{\partial S[h/T,1/T]}{\partial (h/T)}$ we can obtain $\tfrac{h}{T}$ as a function of $M$ and $\tfrac{1}{T}$, and using this solution in Eq.\ \eqref{PMJ} we have $S[M,\frac{1}{T}]$. The detailed expressions are given in Appendix C. We wish to verify Eq.\ \eqref{difPsiex} that, in the present situation, reads
\begin{align}\label{difPMJ}
	\Delta\tilde{S}_\text{ex}[M,\tfrac{1}{T}] &= \tilde{S}_\text{ex}[M,\tfrac{1}{T}] - S_\text{ex}[M,\tfrac{1}{T}] \nonumber \\ 
    &= -\frac{1}{2} \ln \frac{\langle (\Delta \hat{M})^2 \rangle}{\langle (\Delta \hat{M})^2 \rangle_\text{id}},
\end{align}
where $S_\text{ex}[M,\frac{1}{T}] = S[M,\frac{1}{T}] - S_\text{id}(M)$, and $S_\text{id}(M)$ is given by \eqref{Psim2t}. The mean-squared fluctuations are $\langle (\Delta\hat{M})^2\rangle =  -(\frac{\partial^2 S[M,1/T]}{\partial M^2})^{-1}$; the ideal fluctuations were already calculated in \eqref{fluctM}.

\begin{figure}
    \includegraphics[width=\linewidth]{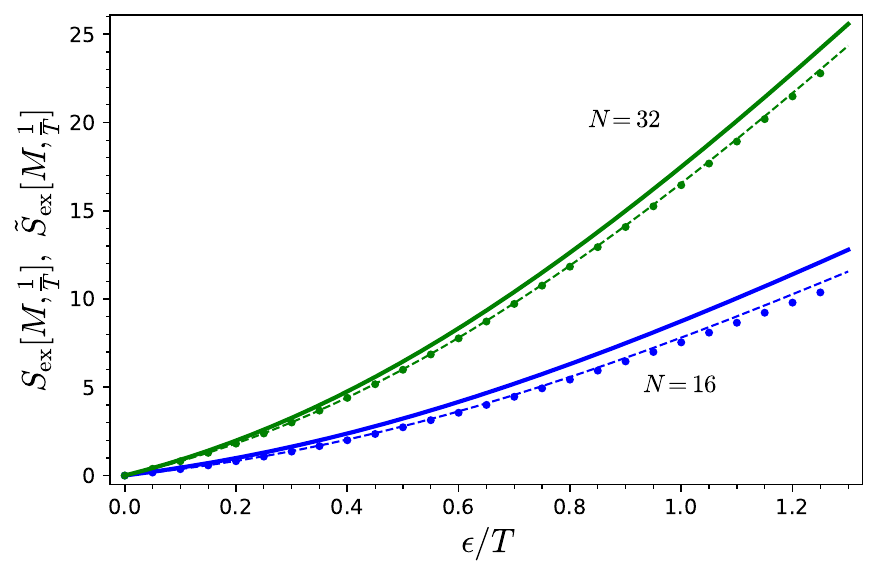}
    \caption{TH Massieu function $S_\text{ex}[M,\tfrac{1}{T}]$ [continuous curves) and SM Massieu function $\tilde{S}_\text{ex}[M,\tfrac{1}{T}]$ (dashed curves, using the correction of Eq.\ \eqref{difPMJ}] against $\epsilon/T$ for the one-dimensional spin lattice. Blue curves are for $N=16$, green curves, for $N=32$. Dots correspond to the numerical evaluation of $\tilde{S}_\text{ex}[M,\tfrac{1}{T}]$; in both cases, an external field is applied such that $M=N/2$.}
    \label{f.spins0}
\end{figure}

\begin{figure}
    \includegraphics[width=\linewidth]{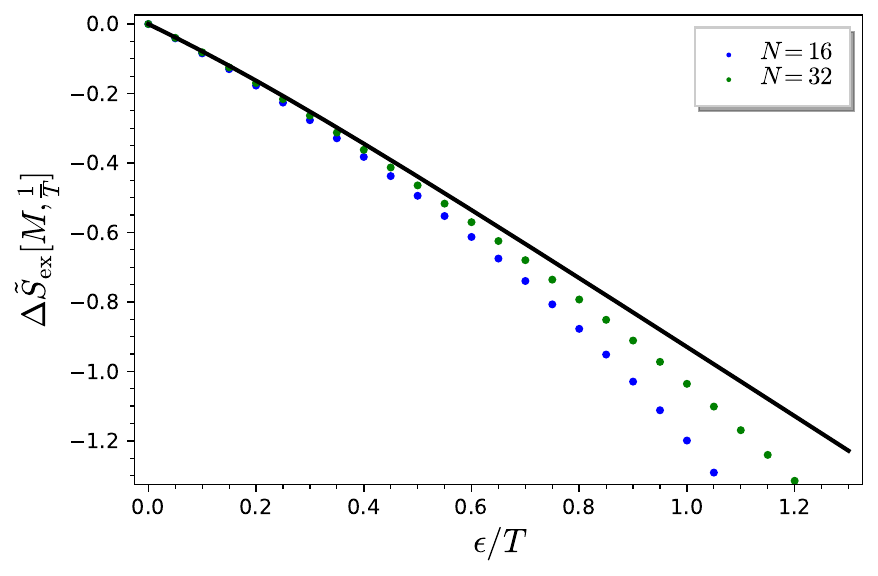}
    \caption{Nonextensive term $\tilde{S}_\text{ex}[M,\tfrac{1}{T}] - S_\text{ex}[M,\tfrac{1}{T}]$ against $\epsilon/T$ for the one-dimensional spin lattice. The curve corresponds to Eq.\ \eqref{difPMJ}. Dots where numerically obtained for $N=16$ and 32; in both cases, an external field is applied such that $M=N/2$. Numerical results approach the theoretical curve as $N$ increases.}
    \label{f.spins}
\end{figure}

We calculate $\tilde{S}_\text{ex}[M,\frac{1}{T}] = \tilde{S}[M,\frac{1}{T}] - \tilde{S}_\text{id}(M)$ numerically. From \eqref{PMT0}, evaluated at $\hat{M}=M$, we have
\begin{equation}\label{PMJ2}
	\tilde{S}[M,\tfrac{1}{T}] = \ln\left( \sum_{\hat{E}} \Omega(M,\hat{E}) e^{-\frac{1}{T}\hat{E}}\right) = \ln \sum_{\omega_M}  e^{-\frac{1}{T}\hat{E}}.
\end{equation}
In order to evaluate the sum in the last equation, we consider a system small enough so that all possible microstates $\omega_M$, for a given magnetization $M$, can be computed. The ideal value, $\tilde{S}_\text{id}(M)$, is given by Eq.\ \eqref{Psim} with $\hat{M}=M$. The numerical results are obtained for values of $\epsilon/T$ between 0 and 1.2, and for $M=N/2$. Figures \ref{f.spins0} and \ref{f.spins} show results for $N$ equal 16 and 32. Figure \ref{f.spins0} shows $S_\text{ex}[M,\tfrac{1}{T}]$ and $\tilde{S}_\text{ex}[M,\tfrac{1}{T}]$ as functions of $\epsilon/T$ with continuous and dashed curves, respectively; the dots represent numerical results for the SM Massieu function $\tilde{S}_\text{ex}[M,\tfrac{1}{T}]$. The significance of the nonextensive term is given by the difference between these curves; the percentage of this correction, for $\epsilon/T=1$, is about 10\% for $N=16$ and 5\% for $N=32$. Figure \ref{f.spins} shows with more detail the accuracy of the theoretical prediction for the difference $\Delta\tilde{S}_\text{ex}[M,\tfrac{1}{T}]$, see Eq.\ \eqref{difPMJ}.  Theoretically, this difference is an intensive quantity independent of the system size $N$, while the numerical results depend on $N$, showing a better agreement with the theory for larger $N$. This is an expected behavior since the theoretical results hold for large $N$. 


\section{Summary and conclusions}
\label{s.conclusions}

In a review on finite system thermodynamics \cite{swendsen}, Swendsen suggested that ``For future work on the properties of small systems, the most promising approach seems to be a straightforward application of statistical mechanics.'' The calculation of nonextensive terms of Massieu functions presented here intends to represent one step in that direction. nonextensive terms order $\ln N$ are just one ingredient to be taken into account in the analysis of small systems; there are other effects not mentioned here that are analyzed in, for example, \cite{hill2,bedeaux}.

The nonextensive term of any Massieu function can be written in terms of the determinant of the fluctuation matrix; see Eq.\ \eqref{difPsis3}. The result was analytically verified in the ideal case, when interactions can be neglected, and numerically verified in the nonideal case. The examples analyzed were particle systems with hard disk and hard sphere interactions and a one-dimensional spin lattice. The procedure to obtain the nonextensive term is based on a Taylor expansion of equilibrium fluctuations where terms of order larger than 2 are neglected; this approximation is justified for large $N$, since the $i$th derivative of a (extensive) Massieu function respect to $i$ extensive variables is $O(N^{-i+1})$ and the central moments are $O(N)$; see Eq.\ \eqref{der2}, or Eq.\ \eqref{avs} for the general derivation. For this reason we expect a better performance of the results for larger values of $N$ and large departures close to a critical point where fluctuations diverge. We observed that the difference between theory and numerical results decreases as $N$ increases, in agreement with this prediction. An increase in $N$ also implies, in our examples, an increase in the system size that reduces the implicit or anomalous finite-size effects caused by the periodic boundary conditions.

Numerical simulations show that there is a regime, characterized by weak interactions (corresponding to small concentration $\rho$ in a particle system, or small coupling $\epsilon$ in a spin lattice), for which the prediction for the nonextensive term holds even for quite small values of $N$. Discrepancies with the theory appear for large concentration in particle systems or for strong interaction in the spin lattice. A criterion to determine whether a system is small or large is to compare the system size with the fluctuation correlation length. Implicit finite-size corrections become relevant if the correlation length is similar to the system size; we consider that this is the cause of the discrepancies that appear for stronger interactions (large densities for the particle systems or large coupling for the spin lattice). This argumentation, based on the correlation length criterion presented in \cite{lebowitz-percus}, is qualitative; further research is needed to quantitatively determine the validity range of the results.

It was shown that the nonextensive term of the excess free energy of a particle system, Eq.\ \eqref{difex3}, is consistent with previous results obtained for the excess chemical potential using the Widom insertion formula \cite{siepmann,dimuro}. There is an important connection between nonextensive terms and transition probabilities that was first explored in Ref.\ \cite{dimuro} in the context of systems that have fluctuations only in the number of particles $N$. This connection will be further explored in future work using the more general results developed here.

\begin{acknowledgments}
	This work was partially supported by Consejo Nacional de Investigaciones Científicas y Técnicas (CONICET, Argentina, PUE 22920200100016CO).
\end{acknowledgments}

\section*{Appendix A}

Equation\ \eqref{difPsis3} can be used to demonstrate that the difference $\Delta\tilde{S}[\vY]$ for the last Legendre transform is a constant. 

We obtain the probability of fluctuations, $\Delta \hat{\vX} = \hat{\vX} - \vX$, directly from the probability of $\hat{\vX}$, Eq.\ \eqref{PX}. For simplicity, we consider here the ensemble where all extensive variables (except volume) have fluctuations ($s=r$); it is described by the first Massieu function, that is, the entropy $\tilde{S}(\hat{\vX})$. Expanding $\tilde{S}(\hat{\vX})$ in a Taylor series up to second order in $\Delta\hat{\vX}$, we have
\begin{align}
    \tilde{S}&(\vX + \Delta\hat{\vX}) \nonumber\\ 
    &=\tilde{S}(\vX) + \frac{\partial \tilde{S}(\vX)}{\partial \vX} \cdot \Delta\hat{\vX} + \frac{1}{2} \Delta\hat{\vX} \cdot \frac{\partial^2 \tilde{S}(\vX)}{\partial \vX^2}\cdot \Delta\hat{\vX}\nonumber \\
    &= \tilde{S}(\vX) + \vY \cdot \Delta\hat{\vX} - \frac{1}{2} \Delta\hat{\vX} \cdot C^{-1} \cdot \Delta\hat{\vX},
\end{align}
where we used that $\frac{\partial \tilde{S}(\vX)}{\partial \vX} = \vY$ and that the inverse of the fluctuation matrix (or covariance matrix) is $C^{-1} = -\frac{\partial \vY}{\partial \vX}$. Replacing in \eqref{PX}, we obtain
\begin{equation}\label{PDX}
	P(\Delta\hat{\vX}) = e^{-\tilde{S}[\vY] + \tilde{S}(\vX) - \vY\cdot\vX - \frac{1}{2} \Delta\hat{\vX} \cdot C^{-1} \cdot \Delta\hat{\vX}}.
\end{equation}
Using the Legendre transform $S[\vY] = S(\vX) - \vY\cdot\vX$, and $\Delta\tilde{S}(\vX) = \tilde{S}(\vX) - S(\vX)$ and $\Delta\tilde{S}[\vY] = \tilde{S}[\vY] - S[\vY]$, we have
\begin{equation}\label{PDX2}
	P(\Delta\hat{\vX}) = e^{-\Delta\tilde{S}[\vY] + \Delta\tilde{S}(\vX)}\,e^{- \frac{1}{2} \Delta\hat{\vX} \cdot C^{-1} \cdot \Delta\hat{\vX}},
\end{equation}
and therefore, the factor $e^{-\Delta\tilde{S}[\vY] + \Delta\tilde{S}(\vX)}$, a function of the equilibrium parameters $\vX$ and $\vY$, is the normalization of the multivariate normal distribution given by $(2\pi)^{-r/2} (\det C)^{-1/2}$, where $r$ is the number of variables. See, for example, \cite{mishin} or Sec.\ 7.C in \cite{reichl}; the multivariate Gaussian distribution is a consequence of the central limit theorem, see Ch.\ V in \cite{khinchin}. 

Then,
\begin{equation}
    e^{-\Delta\tilde{S}[\vY] + \Delta\tilde{S}(\vX)} = (2\pi)^{-r/2} (\det C)^{-1/2},
\end{equation}
or
\begin{equation}
    \Delta\tilde{S}(\vX)-\Delta\tilde{S}[\vY] = -\frac{1}{2}\ln (\det C) -\frac{r}{2}\ln(2\pi).
\end{equation}
Now, using Eq.\ \eqref{difPsis3}, we have that $\Delta\tilde{S}[\vY]$ is a constant:
\begin{equation}
    \Delta\tilde{S}[\vY] = c + \frac{r}{2}\ln(2\pi).
\end{equation}

\section*{Appendix B}

A straightforward method to obtain the average used in Eq.\ \eqref{FexZ} is as follows. In a unit box with periodic boundary conditions we insert $N$ equal particles of diameter $\sigma$ on random positions and then calculate the interaction energy. The pairwise interaction between particles $i$ and $j$ is
\[ \phi_{i,j}=\begin{cases} 
      0 & \|\textbf{r}_i-\textbf{r}_j\| \geq \sigma \\
      \infty & \|\textbf{r}_i-\textbf{r}_j\| < \sigma, 
   \end{cases}
\]
where $\textbf{r}_i$ is the position of the center of particle $i$. Thus, for a given configuration the exponential in Eq.\ (\ref{FexZ}) can be either 0 or 1. The average is then obtained by exploring a large number of particle configurations.
A drawback of this method is that for a large number of particles and moderate concentration values, the probability of randomly obtaining a non-overlapping configuration of particles is vanishingly small. Thus, the number of configurations that need to be explored becomes significantly high.

We employ an alternative approach to tackle this issue. For a hard disk or hard sphere interaction the average in Eq.\ (\ref{FexZ}) is simply the probability of a non-overlapping configuration when the particle positions are chosen at random. Let us call $P(\bar{I}_N)$ the probability that none of the $N$ particles overlap or, in other words, that the intersection among them is empty: $P\left(\bigcap_{i=1}^{N} o_i =\emptyset\right)$, where $o_i$ is the area or volume occupied by particle number $i$. Also, we call $P(\bar{I}_j|\bar{I}_{j-1})$ the conditional probability that $j$ particles do not overlap given that $j-1$ do not, or $P\left(\bigcap_{i=1}^{j} o_i =\emptyset\big|\bigcap_{i=1}^{j-1} o_i =\emptyset\right)$. We can write
\begin{equation}
    P(\bar{I}_N) = P(\bar{I}_2) \prod_{j=3}^N P(\bar{I}_j|\bar{I}_{j-1}).
\end{equation}
We then can calculate the latter probability through the next procedure:
\begin{enumerate}
\item Set the positions of the $j-1$ particles in a close packed arrangement, hexagonal, for instance.
\item Perform a thermalization process in order to reach a typical equilibrium configuration.
\item Insert an additional particle and check whether it overlaps with the rest or not. 
\end{enumerate}

On each time step of the thermalization process, a number $j$ of particles is randomly selected and displace a fixed distance $d$ in a random direction. In case a particle overlaps with any of the rest in the new position, then the particle remains in its original position.
For the numerical simulations we used 1000 time steps for the thermalization process. The displacement $d$ was chosen depending on the concentration value, in order to maximize the particle mixing.

\section*{Appendix C}

The expressions for fluctuations, external field and Massieu functions for the one-dimensional spin lattice are given in this appendix. The field as a function of $M$ and $1/T$ is
\begin{equation}
    \frac{h}{T}=\log\left(A + \sqrt{A^{2} + 1}\right),  
\end{equation}
with $A=M e^{-2 \epsilon/T}/\sqrt{N^{2} -M^{2}}$. 
Using this equation, the TH Massieu function $S[M,\tfrac{1}{T}]$ is
\begin{align}
    S[M,&\tfrac{1}{T}] = \underbrace{N_+ \ln \frac{N}{N_+} + N_- \ln \frac{N}{N_-}}_{\text{ideal  part}} + \frac{\epsilon}{T}(2M - N) \nonumber\\ 
    & - M\ln\left( \tfrac{1 + \sqrt{1+A^{-2}}}{1+N/M} \right) + N \ln\left( \tfrac{N + M \sqrt{1 + A^{-2}}}{2N} \right).
\end{align}
And the mean-squared fluctuations $\langle (\Delta\hat{M})^2\rangle =  -(\frac{\partial^2 S[M,1/T]}{\partial M^2})^{-1}$ are given by
\begin{equation}
    \langle (\Delta \hat{M})^2\rangle = \tfrac{{\left(2 \, A^{3} + 2 \, \sqrt{A^{2} + 1} A^{2} + 2 \, A + \sqrt{A^{2} + 1}\right)} \left(N^2 - M^2\right) M}{{\left(2 \, A^{2} + 2 \, \sqrt{A^{2} + 1} A + 1\right)} A N^{2}}.
\end{equation}

\bibliography{nonext.bib}

\end{document}